\documentclass[
twocolumn,
showpacs,preprintnumbers,amsmath,amssymb]{revtex4}

\usepackage[utf8]{inputenc}
\usepackage{amsmath}
\usepackage{subfigure}
\usepackage{array}
\usepackage{graphicx}
\usepackage{fancyheadings}
\usepackage{dcolumn}% Align table columns on decimal point
\usepackage{bm}% bold math

\def\be{\begin{equation}}
\def\ee{\end{equation}}
\def\ba{\begin{eqnarray}}
\def\ea{\end{eqnarray}}

\begin{document}

\title{A self-consistent renormalized Jellium approach for calculating structural and thermodynamic properties of charge stabilized colloidal suspensions}% Force line breaks with \\

\author{Thiago E. Colla}
\email{colla@if.ufrgs.br}
  \author{Yan Levin}
  \email{levin@if.ufrgs.br}
\affiliation{
Instituto de F\'isica, Universidade Fedaral do Rio Grande do Sul, CP 15051, 91501-970 Porto Alegre, RS, Brazil.
}

\author{Emmanuel Trizac}
\email{trizac@lptms.u-psud.fr}
\affiliation{CNRS, Universit\'e Paris-Sud, UMR 8626, LPTMS, F-91405, Orsay Cedex, France }
\date{\today}% It is always \today, today,
             %  but any date may be explicitl

\begin{abstract}

An approach is proposed which allows to self-consistently calculate the structural and thermodynamic properties of highly charged  aqueous colloidal
suspensions.  The method is based on the renormalized Jellium model with the background charge distribution
related to the colloid-colloid correlation function.  The theory is used to calculate the correlation functions
and the effective colloidal charges for suspension containing additional monovalent electrolyte.  The predictions
of the theory are in excellent agreement with the Monte Carlo simulations.
\end{abstract}

\maketitle
\section{Introduction}

Over the span of the last century, colloidal suspensions have been the subject of intense
theoretical and experimental study. The great effort is well justified
by the importance that these systems play
in industrial, biological, and medical applications.  A practical problem
that arises is how to stabilize suspensions against flocculation and precipitation, resulting 
from short range attractive van der Waals interaction.  
One approach is to synthesize colloidal particles with acidic groups on their surface.
In aqueous environment these groups become ionized, resulting in repulsion between the macroions.

Charge stabilized colloidal suspensions are an extreme example of a large asymmetry electrolyte.  Both the charge and the size of the macroions are orders of magnitudes larger than those of other ionic species present inside the suspension. Typically a colloidal particle of radius $1000$ \AA, will carry $10^{3}-10^{4}$ ionizable groups uniformly distributed over its surface. The huge asymmetry between the macroions and the microions makes the theoretical investigation of colloidal suspensions a very difficult task \cite{1,2,2a,2b}. The standard approach used to study these systems 
is based on the Primitive Model (PM), which treats solvent as a dielectric continuum of permittivity $\epsilon$. The interaction potential between the ionic species is
taken to be composed of a long range Coulomb interaction and a short range hard-core repulsion. Unfortunately, due to the large charge and size asymmetry between the
macroions and the microions, even for this simplified model the traditional methods of liquid state theory --- such as the molecular dynamics simulations, the Monte Carlo (MC) simulations, and the integral equations theories --- prove to be only partially successful~\cite{2}. The huge number of counterions needed to ensure the bulk electroneutrality, allows the maximum charge asymmetry which can be studied using the present day computers to be around  $100:1$, 
while the important case of added electrolyte remains practically unaccessible. 
Similarly, integral equation theories are plagued by convergence problems for strongly asymmetric electrolytes~\cite{3}.

To obtain a more tractable description of these systems it is, therefore, necessary to introduce further simplifications.  
This can be achieved by integrating out the microion degrees of freedom, leaving only a
state dependent interaction potential between the colloidal particles.  This defines the, so called, one component model (OCM). In spite of its apparent simplicity, the OCM requires knowledge of the effective macroion-macroion interaction, which implicitly depends on all the ionic species. Formally, the potential can be obtained by explicitly tracing out the degrees of freedom of the microions of the PM~\cite{2,4}. In practice, however, this coarse graining procedure can only be accomplished by means of approximate theories, such as the Poisson Boltzmann \cite{4,5}  or the Ornstein-Zernike (OZ) equations, with appropriate closure relations \cite{3,6,7}. Furthermore, to avoid computational difficulties one usually assumes that the effective interaction potential is pairwise additive. This is quite reasonable at low macroion concentrations, however, care must be used when applying this assumption to more concentrated systems. As the concentration increases, the many-body correlations start to play an important role for both structural and thermodynamic properties. 
Assuming the OCM description with pairwise macroion interactions, there still remains a question of how to obtain the effective interaction potential. This has been the subject of many works \cite{1,8,9,10}. The difficulty in answering it is due to various factors, among which are strong correlations between the various particles and the huge asymmetry between the different ionic species ---  forcing different approximations for different correlations. In the limit of large dilutions and small colloidal charge, a linearized Debye-H\"uckel theory can be applied, and the pair potential takes a simple Yukawa-like form,  known as the Derjaguin-Landau-Overbeek-Verwey (DLVO) potential. For a system of colloidal particles of radius $a$, charge $-Z_{o} q$, density $\rho_{o}$, and  microions of valence $z_{i}$ and bulk concentrations $\rho_{i}$, $i>0$, the DLVO potential is given by:
\begin{equation}
 \beta u(r)= \ell_{b}\left({\dfrac{Z_{o}}{(1+\kappa a)}}\right)^{2}\dfrac{e^{-\kappa (r-2a)}}{r},
\end{equation}
where $\ell_{b}=\beta q^{2}/\epsilon$ is the Bjerrum length, $\beta=1/k_{B}T$, $q$ is the elementary charge, $\epsilon$ is the dielectric constant, and $\kappa=\sqrt{4\pi\ell_{b}\sum_{i\neq o}\rho_{i}z_{i}^{2}}$ defines the inverse Debye screening length. From Eq. (1) one can see that,  
at this level of approximation, the role of the small ions is only to screen the electrostatic interaction between the macroions \cite{1}.

Even though the potential in Eq. (1) is restricted to low concentrations and small colloidal charges, the  {\it functional form} of the DLVO potential can be extended to describe systems at moderate and high concentrations as well as large colloidal charge. To do this the structural charge $Z_{o}$ is replaced by an effective charge ${Z}_{eff}$, which accounts for the non-linear effects of counterion condensation~\cite{1,3,11,12}. In fact, it can be formally shown that non-linear short range correlations within the PM electrolytes can all be introduced into the DH theory by means of appropriate renormalization procedures \cite{13}. The physical picture behind the charge renormalization is that strong electrostatic attraction between the macroions and the counterions leads to
their association, so that from large distances (compared to the Debye length), a macroion can be viewed as carrying charge  smaller than its structural bare charge. Both the macroion and its layer of condensed counterions can then be considered as forming a single entity
of effective charge ${Z}_{eff}$. Once the non-linear correlations are taken into account through the charge renormalization, 
the DLVO pair potential, Eq. (1), can be used in the OCM description to account for structural properties of colloidal suspensions.

In this paper, we propose an ansatz which allows us to calculate both the thermodynamic and structural properties of charge stabilized colloidal suspensions in a fully self-consistent way. This ansatz is based on a coupling of the renormalized Jellium model~\cite{14,15} with the OCM Ornstein-Zernike integral equations theory.  From now on, we will only consider the case of aqueous monovalent electrolytes ($z_{j}=z_{\pm}=\pm 1$).

\section{Theoretical background}

Most of the theoretical work to obtain the effective charge of colloidal particles is based on the mean field Poisson-Boltzmann equation \cite{11}. In many cases, the infinite dilution limit is employed, and the problem reduces to that of a spherical macroion or an infinite planar wall immersed in electrolyte. For a more realistic situation of finite macroion concentration, the colloidal distribution must be incorporated into the PB equation. To do this, one must solve the PB equation for a fixed macroion configuration  from which the stress tensor 
and the force acting on each macroparticle can be calculated. Clearly, a numerical implementation of such procedure is very difficult~\cite{5,16}. To have
a more tractable approach,
further simplifications are necessary.  In this respect two approximations have proven to be particularly useful: the cell and the 
renormalized Jellium models. Before introducing the new theory, we will make a brief review of the basic features of 
these approximations and discuss how the effective charges can be extracted from them.

\subsection{Renormalization models}

In the cell model, colloids are assumed to have a quasi solid-state like structure --- macroions arranged in a form of a lattice. 
This allows us to consider one macroion in a corresponding Wigner-Seitz (WS) cell. A further approximation is to 
replace the polyhedral WS cell by a cell having the same symmetry as the macroion. The size of the cell is obtained from the overall macroion concentration. 
Because of the charge neutrality, the electric field must vanish on the surface of each cell, so that within this approach there is no pair interaction
between the colloidal particles. Nevertheless, the model is often used to calculate the effective macroion charges, which enters into
the DLVO pair interaction potential. To obtain the effective charge, the  non-linear PB equation is solved numerically inside the WS cell. The solution is then
asymptotically matched to that of the {\it linearized} PB equation with an {\it effective} charge --- the so called Alexander prescription \cite{17,18}. 

The Jellium model captures the opposite limit in which the colloid-colloid correlation function is assumed to be completely 
disordered, $g_{oo}(r) \approx 1$  \cite{7}. This approach is well suited for low density, weakly charged colloidal particles. For strongly
charged macroions, the Jellium approximation fails to converge.
Recently, Trizac and Levin have proposed a renormalization procedure designed to extend the validity of the Jellium approximation for strongly charged colloidal particles~\cite{1,2}. The renormalized Jellium model relies on the concept of counterion condensation to determine the effective charge of the macroions. The method works as follows. One macroion with a charge $Z_{o}$ is positioned at the origin of the coordinate system, the remaining macroions with their condensed counterions are assumed to form a uniform neutralizing background in which the uncondensed counterions and coions move freely. 
Because it is not know how many counterions will condense onto the colloidal particles, the background charge density is not know {\it a priori}, but must
be determined self-consistently. The distribution of uncondensed counterions and coion around the central macroion is assumed to be of the Boltzmann form, $g_{oj}(r)=e^{-\beta q z_{j} \psi(r)}$ with $z_{j}=\pm 1$, where $\psi(r)$ is the mean electrostatic potential around the central macroion.  The electrostatic potential satisfies the modified Poisson-Boltzmann equation,
\begin{equation}
 \nabla^{2}\psi(r)=-\frac{4\pi q}{\epsilon}\left(\sum_{j=\pm}\rho_{j}z_{j}e^{-\beta e z_{j} \psi(r)}-Z_{back}\rho_{back}(r)\right),
\label{eq1}
\end{equation}
where $Z_{back}$ and $\rho_{back}(r)$ are the background charge and density, respectively.  
In the canonical ensemble --- fixed number of all particles --- $\rho_j$'s are determined from the overall electroneutrality, while in the  
the semi-grand canonical ensemble~\cite{15}, when the suspension is in contact with a salt reservoir at concentration $c_s$, 
$\rho_j=c_s$.
We note that Eq.(\ref{eq1}) would be exact if the electrostatic potential $\psi(r)$ on the right hand side of Eq.(\ref{eq1}) is replaced by the potential of mean force
between the microion and colloid, $w(r)$.  In that case $Z_{back}$ would simply be the bare colloidal 
charge $Z_o$ and $\rho_{back}(r)=\rho_o g_{oo}(r)$, where $\rho_o$ is the mean colloidal density.
Unfortunately, there is no explicit way of calculating the potential of mean force.  We are, thus, forced to identify 
$w(r) \approx \psi(r)$.  This is permissible for monovalent ions in aqueous suspensions for which the electrostatic correlations between the 
microions are small.
The price for identifying $w(r) \approx \psi(r)$ is, however, a mandatory renormalization of the colloidal charge. Furthermore, one looses
the direct identity between the background density and the colloid-colloid correlation function.

Within the renormalized Jellium approximation $\rho_{back}(r)=\rho_o$, and the bulk electroneutrality condition becomes
\begin{equation}
 \sum_{j=\pm}\rho_{j}z_{j} e^{-\beta q z_j \psi_\infty}-Z_{back}\rho_o=0.
\end{equation}
where $\psi_{\infty}$ is the Donnan potential which ensures the overall electroneutrality. In the canonical
ensemble, we can take $\psi_\infty=0$.
For a given set of parameters (including the background charge $Z_{back}$), equation (\ref{eq1}) can be solved numerically.  Asymptotically,
its solution has the form 
\begin{equation}
\psi_{as}(r)= \psi_{\infty}-\dfrac{Z_{eff} q}{\epsilon }\frac{e^{-\kappa (r-a)}}{r(1+\kappa a)}. 
\label{eq2}
\end{equation}  
In the semi-grand canonical ensemble~\cite{15} $\beta q \psi_\infty=- \operatorname{arcsinh}(Z_{eff} \rho_o/2 c_s)$ and the 
inverse Debye length is $\kappa=\sqrt{8 \pi\ell_{b} c_s \cosh (\beta q  \psi_\infty)}$. In the canonical ensemble, $\psi_\infty = 0$ and 
$\kappa=\sqrt{4\pi\ell_{b} (\rho_-+\rho_+)}$.  Eq. (\ref{eq2}) allows us to calculate the effective charge $Z_{eff}$
as a function of $Z_{o}$ and $Z_{back}$.  The self consistency condition is imposed by requiring that $Z_{eff}=Z_{back}$, which determines the physical value of the effective colloidal charge.  It is
important to note that unlike the cell model for which there is no pairwise interaction between the colloids, 
the macroion-macroion potential of the renormalized Jellium model is precisely of the DLVO form.

To extend the renormalized Jellium model to larger concentrations,  Casta\~neda-Priego \textit{et al.}~\cite{19} have proposed to modify the uniform
background density $\rho_{back}(r)=\rho_o$, to account for the correlation hole around each macroions.  These authors
observed that for salt-free suspensions, simulations find that the colloid-colloid correlation function has the first maximum at $r\approx \rho^{-1/3}$.  They
then suggested that this distance can be used to fix the size of the correlation hole between the macroions in salt-free suspensions~\cite{19}. 
Casta\~neda-Priego \textit{et al.} suggested that around each macroion there is an effective exclusion zone of radius $r_h=1/2 \rho^{1/3}$, 
devoid of the background charge. 
The factor of two is included in order to account for the fact that the exclusion zone is divided equally between the two macroions, see Fig. 1. 
The exclusion zone around each colloid is then taken into account by replacing the usual uniform Jellium background density by a step 
function $\rho_{back}(r)=\rho_o \Theta(r-r_h)$ in Eq. (2).  
Such procedure, however, still lacks the self-consistency, since the resulting effective charge can not be directly related with the correlation
function, which is implicit in the form of $\rho_{back}(r)$. Furthermore, it is not clear how one can extend the above procedure to define the radius of 
the correlation hole for suspensions containing additional 1:1 electrolyte.

\section{The model}

Although the renormalized Jellium model and its modified versions allow us to calculate the effective charges, both theories lack in internal self consistency.  In order to avoid this, it is necessary to find a way of calculate the effective charge and the correlation function $g_{oo}(r)$ simultaneously. 
To achieve this, we  observe that the background charge in Eq. (\ref{eq1}) should be related in some way to the colloid-colloid correlation
function.  Unfortunately, as discussed above, within the renormalized Jellium model one can not identify the spatial variation of the background charge directly with the $g_{oo}(r)$. We note, however, that $g_{oo}(r)$ does carry the information about the size of the exclusion zone, which is approximately half the distance to the first peak of $g_{oo}(r)$. In view of this observation, we will make the ansatz of identifying the background density variation with
the rescaled colloid-colloid correlation function  $\rho_{back}(r)=\rho_o g_{oo}(2r)$. This choice leads to a uniform background far from colloid $\rho_{back}(r) \approx \rho_o$,  while at the same time produces a correlation hole of appropriate size, $\rho_{back}(r)\approx 0$ for $r<r_h$.

\begin{figure}
\centering
 \includegraphics[width=5cm]{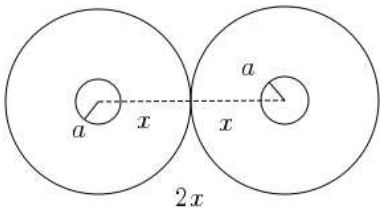}
\caption{Two macroions of radius $a$  separated by a distance $r=2x$, at which the macroion-macroion correlation function has the first maximum. By symmetry, we can define around each macroion an effective exclusion zone of radius $x$.}
\end{figure}

With this modification, the fully self-consistent Jellium equation becomes
\begin{equation}
 \nabla^{2}\phi(\tilde{r})=-4\pi\ell_{b}a^{2}\left(\rho_{+}e^{-\displaystyle\phi(\tilde{r})}-\rho_{-}e^{\displaystyle\phi(\tilde{r})}\right)+3\eta\tilde{Z}_{back}g_{oo}(2\tilde{r}),
\label{eq3}
\end{equation}
where we have defined the dimensionless quantities  $\tilde{r}\equiv r/a$, $\phi(r)\equiv\beta e \psi(r)$, $\tilde{Z}_{back}\equiv Z_{back}\ell_{b}/a$ and $\eta\equiv4\pi a^{3}\rho_o/3$. 
As before, the effective charge is determined by the requirement that $Z_{eff}=Z_{back}$. 
Eq. (\ref{eq3}) is solved by an iterative procedure. We start with colloid-colloid pair correlation $g_{oo}^{(1)}(r)=1$, and use this in Eq. ({\ref{eq3}}) to extract the corresponding effective and background charge $Z_{eff}^{(1)}=Z_{back}^{(1)}$. The system is then considered in the OCM approach, with the interaction potential between the colloids given by,
\begin{equation}
 \beta u(\tilde{r})=\dfrac{1}{\Gamma}\left(\dfrac{\tilde{Z}_{eff}}{(1+\kappa a)}\right)^{2}\dfrac{e^{-\displaystyle{\kappa a (\tilde{r}-2})}}{\tilde{r}},
\label{eq4}
\end{equation}
where $\Gamma=\ell_{b}/a$. In the semi-grand canonical ensemble, $(\kappa a)^{2}= (\kappa_{1}a)^{2}(1+(\kappa_{res}/\kappa_{1})^{4})^{1/2}$,  $(\kappa_{1}a)^{2}=3\eta \tilde{Z}_{eff}$ and $(\kappa_{res}a)^{2}=8\pi\ell_{b}c_{s} a^2$.  
We then numerically solve the one component OZ integral equation with the Roger-Young (RY) closure to determine the new pair correlation function $g_{oo}^{(2)}(r)$. This function is then used as a new input in Eq. (\ref{eq3}) to calculate the new effective charge $Z_{eff}^{(2)}$. The procedure is iterated until the convergence is achieved,  $g_{oo}^{(i)}(r)=g_{oo}^{(i-1)}(r)$. In practice, only a few iterations are necessary to fulfill this condition.

The Roger-Young closure is an interpolation between the hypernetted chain approximation (HNC) and the Percus-Yevick (PY) relation, with an adjustable parameter $\alpha$ chosen 
so as to satisfy the thermodynamic self-consistency in the calculation of the isothermal compressibility \cite{20}. In the salt-free case, the major contribution to the osmotic pressure comes from counterions, so that $\alpha$  is determined by imposing the requirement that
\begin{equation}
 \dfrac{\partial(\beta P)}{\partial\rho}=1+\rho \hat{h}_{oo}(0)\approx \dfrac{\tilde{Z}_{eff}}{\Gamma},
\label{eq5}
\end{equation}
where $\hat{h}_{oo}$ is the Fourier transform of the total correlation function. The first equality is the Kirkwood-Buff relation \cite{21}, while the second one comes from approximating the microion pressure by the Jellium equation of state, $\beta P= Z_{eff}\rho$, and disregarding the weak dependence of 
the effective charge on the macroion density.

For the case of large salt concentrations and moderate volume fractions --- when the density dependence of the effective pair potential is weak --- the pressure is given by that of the OCM~\cite{22,23}, and the last equality in Eq. (\ref{eq5}) is replaced by the OCM inverse compressibility,
\begin{equation}
1+\rho\hat{h}_{oo}(0)=\dfrac{\partial(\beta P_{OCM})}{\partial \rho}.
\label{eq6}
\end{equation}
The OCM pressure $P_{OCM}$ can be calculated from the pair correlation function using the well known virial equation. It is important to note that in calculating the right hand side of Eq.(\ref{eq6}), the interaction potential must be kept constant \cite{22}. We use Eq.(\ref{eq6}) to determine $\alpha$ in the RY closure  when dealing with suspensions in contact with a salt reservoir at large concentration. 

In practice, due to finite discretization, the correlation function does not saturate at unity for large distances. Instead, it oscillates around $1$ with a small amplitude.  This creates difficulty for the solution of the PB equation. In order to ensure the correct long-distance behavior of $\psi(r)$, 
we  set a cut-off distance $\tilde{r}_{c}$, beyond which we force $g_{oo}(\tilde{r})=1$. The value of $\tilde{r}_{c}$ is chosen such that  
$\vert g_{oo}(\tilde{r})-1\vert < 0.0025$ for $\tilde{r}>\tilde{r}_{c}$.

\section{Results}
In Fig. 2 we plot the macroion-macroion correlation functions calculated using the fully 
self-consistent Jellium (sc-Jellium) model developed above, and compare it with the  
 results of the modified
Jellium approximation of Casta\~neda-Priego \textit{et al.}  (m-Jellium) and with the MC simulations performed by Linse \cite{24} for aqueous deionized suspension, $\ell_{b}\approx7.2$\AA, $\kappa_{res}=0$ and colloidal volume fraction $\eta=0.01$. Three systems with coupling parameters $\Gamma=0.3558$ (a), $0.1779$ (b), and $0.0445$ (c), corresponding to particles of radius $a\approx160$\AA, $40$\AA, and $20$\AA, respectively, were studied.  All the
calculations were performed in the saturation limit (very large bare charge). 
The corresponding effective charges are displayed in the graphs. As can be seen, both the sc-Jellium approach and  m-Jellium  show  good agreement with the MC simulations for $\Gamma=0.3558$ and $\Gamma=0.1779$, while for the  lowest value $\Gamma=0.0445$ the m-Jellium model seems to strongly overestimate the colloidal structure.

\begin{figure}[ht]
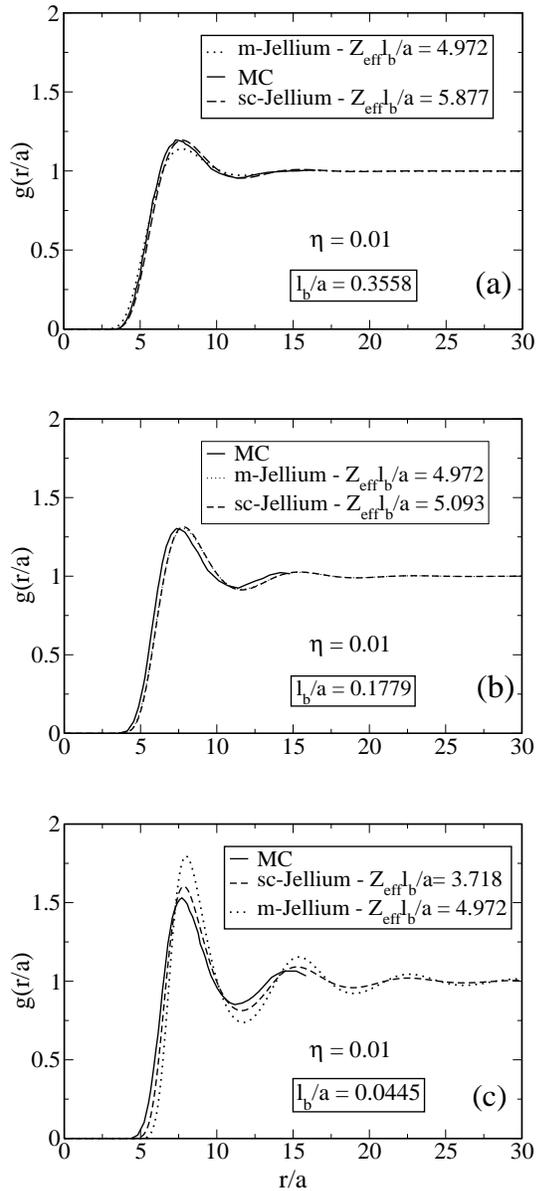

\centering

\subfigure
{\includegraphics[width=7cm,height=5.0cm]{g2r_0.3558_0.01.eps}}

\subfigure{\includegraphics[width=7cm,height=5.0cm]{g2r_0.1779_0.01.eps}}

\subfigure{\includegraphics[width=7cm,height=5.0cm]{g2r_0.0445_0.01.eps}}

\caption{Macroion-macroion correlation functions calculated using the sc-Jellium (dashed lines), m-Jellium (dotted lines) and MC simulations (solid lines) for a deionized colloidal suspensions at volume fraction $\eta=0.01$. The coupling parameters are (a) $\Gamma=0.3558$ , (b) $\Gamma=0.1779$  and (c) $\Gamma=0.0445$. }
\end{figure}

In Fig. 3 we compare the correlation functions calculated using the sc-Jellium with the MC simulations for various colloidal volume fractions at fixed coupling parameter $\Gamma=0.3558$ in the no-salt regime. Again, a good agreement with the MC simulations is found, for all the macroion concentrations. Surprisingly, this agreement seems to be better at higher volume fractions, diminishing as the concentration becomes very low. Figure (4) shows the behavior of the effective charge as a function of the colloidal volume fraction for the sc-Jellium (dashed curve), m-Jellium (dotted curve) and the original 
renormalized Jellium model of Trizac and Levin (solid curve). Although the
qualitative behavior is the same for all three models, there is a significant quantitative variation in the value of the effective charge. The effective charges
predicted by the sc-Jellium lie between those of the m-Jellium and the renormalized Jellium models. 
\vspace{1cm}

\begin{figure}[ht]
 \centering
 \includegraphics[width=7cm,height=5.0cm]{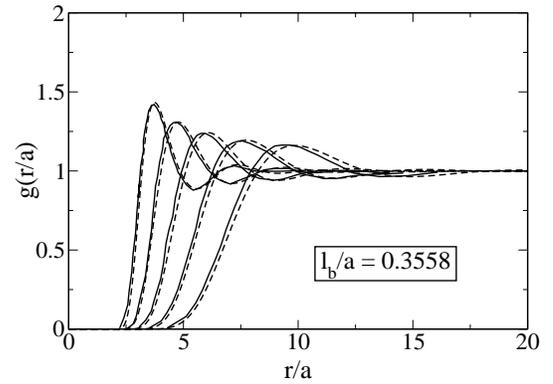}
 % gr_jell.eps: 136560832x136560832 pixel, 300dpi, 1156215.00x1156215.00 cm, bb=
 \caption{Macroion-macroion correlation functions calculated using the sc-Jellium (dashed lines), and MC simulations (solid lines) for a deionized colloidal suspension with coupling parameter $\Gamma=0.3558$. From left to right, the volume fractions are given be $\eta=0.08$, $\eta=0.04$, $\eta=0.02$, $\eta=0.01$, 
and $\eta=0.005$.
\vspace{0.6cm}}
\end{figure}

\begin{figure}[ht]
 \centering
 \includegraphics[width=7cm,height=5cm]{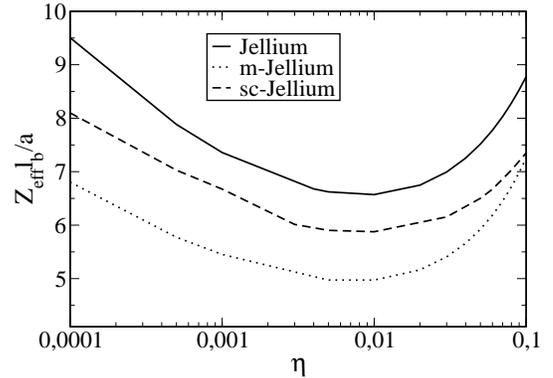}
 % gr_jell.eps: 136560832x136560832 pixel, 300dpi, 1156215.00x1156215.00 cm, bb=
 \caption{Reduced effective charge as a function of volume fraction for a deionized colloidal suspension with $\Gamma=0.3558$, as predicted by the renormalized Jellium model (solid curve), m-Jellium  (dotted curve) and the fully self-consistent approach developed in this paper (dashed curve).}
\end{figure}
The real advantage of the sc-Jellium over the m-Jellium is that it allows us to accurately calculate the effective charges and structures for suspensions
containing 1:1 electrolyte.  At the moment this is the only theory which is capable of doing this for strongly charged colloidal particles. 
The effects of non zero salt concentration on the macroion structure can be seen in Fig. (5), where the correlation functions for
reservoir salt concentrations corresponding to $\kappa_{res}a=1.0$ and $\kappa_{res}a=1.5$ are displayed for various volume fractions. 
Unfortunately, because of the  difficulty of simulating these systems, no MC data is available.  
We see some very general trends for suspensions containing 1:1 electrolyte. As expected, increase of salt concentration leads to 
larger screening and loss of colloidal structure.  In the salt dominated regime, the correlation functions become nearly independent of the macroion 
concentration.  For these cases, both the effective potential and the effective colloidal charge show 
very slow variation with the colloidal volume fraction --- this also explains the weak variation of the correlation functions. 
Another remarkable feature is that at high salt concentrations, 
the colloidal structure is no longer important for the computation of the effective charge,
Fig. (6).  The effective charge calculated using the 
the original renormalized Jellium model with a uniform background and  
the sc-Jellium are practically the same. 
\vspace{0.6cm}
\begin{figure}[ht]
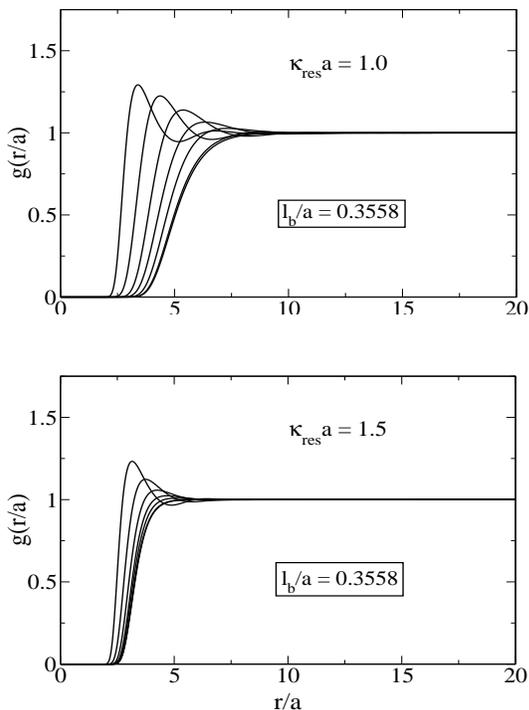

\centering

\subfigure
{\includegraphics[width=7cm,height=4.5cm]{g2r_1.0.eps}}

\subfigure{\includegraphics[width=7cm,height=4.5cm]{g2r_1.5.eps}}
\caption{Macroion-macroion correlation functions calculated using the self-consistent approach, for suspension with $\Gamma=0.3558$ in contact with a salt reservoir at $\kappa_{res}a=1.0$ in (a) and  $\kappa_{res}a=1.5$ in (b). From left to right, colloidal volume fractions are $\eta=0.01$, $\eta=0.02$, $\eta=0.04$, $\eta=0.01$, $\eta=0.005$, $\eta=0.001$ and  $\eta=0.0005$. As the volume fraction decreases, the correlation functions saturate, and the curves start to overlap.
\vspace{0.5cm}}

\end{figure}

\vspace{0.8cm}
\begin{figure}[ht]
 \centering
 \includegraphics[width=7cm,height=5cm]{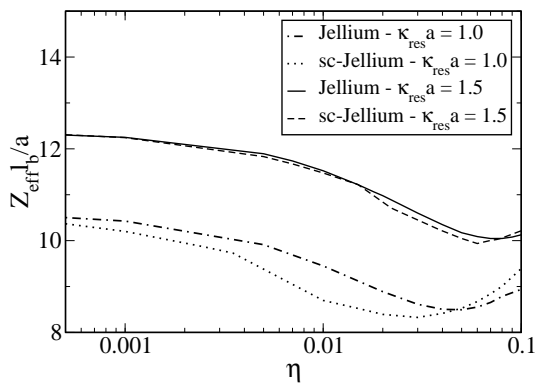}
 % gr_jell.eps: 136560832x136560832 pixel, 300dpi, 1156215.00x1156215.00 cm, bb=
 \caption{Reduced effective charge as a function of volume fraction as predicted by: self-consistent approach with $\kappa_{res}=1.5$ (dashed curve), and $\kappa_{res}=1.0$ (dotted curve)  }
\end{figure}

\section{Summary and conclusions}

We have developed a theory which allows us to self-consistently calculate both the thermodynamic and structural properties of aqueous colloidal
suspensions containing 1:1 electrolyte. The theory is based on coupling the OZ equation with RY closure to the PB equation with the
renormalized Jellium approximation.  
For salt free suspensions, the predictions of the theory were compared to the  MC simulations and were found to be in excellent agreement.
The theory was then applied to study colloidal structure in suspensions containing 1:1 electrolyte.  Unfortunately, no simulational
data is available for these systems.  Nevertheless, we expect that due to its internal self-consistency, the approach developed in this
paper will remain very accurate for these systems as well.  

\section{Acknowledgments}
T.E.C. would like to acknowledge useful conversations with S. Pianegonda during the early stages of this work.
This research is supported in part by CNPq, INCT-FCx of Brazil, and by the Air Force Office of Scientific Research 
(AFOSR), USA, under the grant FA9550-09-1-0283.

\end{document}